\documentstyle[twocolumn,aps,epsf]{revtex}
\begin{document}
\draft
\preprint{IUCM97-037}
\title{Shadow Bands and Tunneling Magnetoresistance in 
Itinerant Electron Ferromagnets}
\author{A.H. MacDonald$^{1}$, T. Jungwirth$^{1,2}$, and M. Kasner$^{1,3}$}
\address{$^{1}$Department of Physics, Indiana University, Bloomington IN 47405}
\address{$^{2}$Institute of Physics ASCR,
Cukrovarnick\'a 10, 162 00 Praha 6, Czech Republic}
\address{$^{3}$Institut f\"ur Theoretische Physik,
Otto-von-Guericke-Universit\"at,PF 4120, D-39016 Magdeburg, Germany}
\date{\today}
\maketitle

{\tightenlines
\begin{abstract}

In itinerant electron ferromagnets spectral weight 
is transferred at finite temperatures from quasiparticle peaks 
located at majority and minority-spin band energies to 
shadow-band peaks.  For a given Bloch wavevector and band index,
the majority-spin shadow-band peak is located near the minority-spin 
quasiparticle energy and the minority-spin shadow-band peak is located 
near the majority-spin quasiparticle energy.  This property 
can explain much of the temperature dependence seen in the 
magnetoresistance of magnetic tunnel junctions.  

\end{abstract}
}


\narrowtext

Efforts\cite{review} to achieve a complete understanding of 
late transition metal ferromagnets have been frustrated 
by fundamental difficulties associated with the band character
of the electrons which carry the spontaneous magnetic moment.
The itinerant character of magnetic electrons in these 
systems is incontrovertibly established by 
de Haas-van Alphen studies\cite{dHvArefs} which map out majority
and minority spin Fermi surfaces enclosing 
$\vec{k}$-space volumes consistent with the measured saturation moment per atom.
Nevertheless, many finite-temperature properties of these systems 
are inconsistent with a simple Stoner-Wolfarth\cite{stoner}
mean-field theory for band ferromagnets and are more easily
rationalized in a picture in which the $d$-electrons are regarded as localized.
The localized {\em vs.} itinerant conundrum is often most 
informatively addressed by probes of the one-particle Green's
function. For example, photoemission studies have\cite{photoold}
and continue\cite{photonew} to improve insight.  The present work is motivated 
by recent success\cite{mtjexpt}
in fabricating ferromagnet-insulator-ferromagnet magnetic tunnel junctions
(MTJ's) with reproducible characteristics, opening up the possibility of 
obtaining spin-resolved information on the tunneling density-of-states
of band ferromagnets.  We point out that, in an itinerant
electron ferromagnet, a portion of the 
spectral weight is transferred at finite temperatures from 
a majority or minority spin quasiparticle peak to a 
shadow-band peak located near the opposite spin quasiparticle energy.
The fraction of the spectral weight transferred is 
proportional, at low temperatures, to the saturation moment 
suppression.  We propose that this effect is responsible for 
much\cite{othereffects} of the temperature-dependence of
MTJ magnetoresistance (MR). 

Tunneling measurements are usually interpreted by 
assuming a weak link which can be modeled by a phenomenological
tunneling Hamiltonian.  This approach allows the tunneling current
to be expressed quite generally in terms of electronic
spectral functions\cite{wolf}.
Assuming only that the temperature is much smaller
than the respective band widths, the tunneling conductance between 
two weakly linked ferromagnets is
\begin{eqnarray}
G &=& \frac{2 \pi e^2}{\hbar} \sum_{\sigma} \sum_{n_L,\vec{k}_L}
\sum_{n_R,\vec{k}_R} |t(n_L,\vec{k}_L;n_R,\vec{k}_R) |^2 \times 
\nonumber \\
& &
A^{L}_{n_L,\vec{k}_L,\sigma}(E_F) A^{R}_{n_R,\vec{k}_R,\sigma}(E_F). 
\label{tuncond}
\end{eqnarray} 
Here $A_{n,\vec{k}}(E)$ is the spectral weight function for band $n$ at
wavevector $\vec{k}$, $L$ (left) and $R$ (right) label
ferromagnets on opposite sides of the MTJ.
and $\sigma=\uparrow$ (majority spin on left of junction ), $(\downarrow)$ 
(minority spin on left side of junction) labels spin.
To establish notation we first discuss our view of the  
band theory interpretation of MTJ MR.

In band theory, metallic ferromagnets are 
characterized by spin-split temperature-independent energy bands with  
infinite quasiparticle lifetimes:
$A_{n,\vec{k},\sigma}(E) = \delta (E_{n,\vec{k},\sigma}-E)$. 
We assume that the tunneling amplitudes appearing in 
Eq.(~\ref{tuncond}) can be approximately decoupled into
factors depending separately on band wavefunctions on
opposite sides of the barrier, {\em i.e.} that 
$|t(n_L,\vec{k}_L;n_R,\vec{k}_R) |^2  \approx 
|t(n_L,\vec{k}_L)||t(n_R,\vec{k}_R)|$.  This assumption is  
physically natural and, as we comment below, is necessary to explain the 
common success\cite{moodera} of the Julli{\` e}re\cite{julliere} formula
in interpreting MR data.
The tunneling conductance is then proportional to a sum over 
spin directions of the product of factors depending separately on 
left and right ferromagnets:  
\begin{equation} 
G = \frac{ 2 \pi e^2}{\hbar} \sum_{\sigma}  t^{L}_{\sigma}
N^{L}_{\sigma} t^{R}_{\sigma} N^{R}_{\sigma}. 
\label{tuncondb}
\end{equation}
Here $N_{\sigma}$ is the Fermi level density of states and 
$t_{\sigma}$ is a weighted average of tunneling amplitude
factors defined by
\begin{equation}
t_{\sigma} N_{\sigma} = \sum_{n,\vec{k}} |t_{n,\vec{k}}|
\delta(E_F - E_{n,\vec{k},\sigma}).
\label{tunampdef}
\end{equation}

{}From Eq.(~\ref{tuncondb}) 
we can evaluate the parameter usually
used to characterize the size of the magnetoresistance of a 
MTJ:
\begin{equation}
MR\equiv
\frac{G_{P}-G_{A}}{G_{P}} = \frac{ 2 P^{L} P^{R}}{ 1 + P^{L} P^{R}} \, . 
\label{julliere}
\end{equation}
Here $G_{P}$ and $G_{A}$ are respectively the conductances when  
ordered moments on opposite sides of the junction have parallel
and antiparallel orientations. 
\begin{equation}
P = \frac{t_{a} N_{a} - t_{i} N_{i}}
{ t_{a} N_{a} + t_{i} N_{i}}
\label{tuncurpolar}
\end{equation}   
where the indices $a$ and $i$ refer to m{\it a}jority and 
m{\it i}nority spins respectively.
Comparing Eq.(~\ref{tuncurpolar}) and Eq.(~\ref{tuncondb}), $P$ can
be identified as the spin polarization of the tunneling current 
between a ferromagnet and a spin-unpolarized system.  This quantity has
been measured for most systems of interest\cite{tedrow}. 
The tunneling current is generally found to be dominated by
the majority spins, even when their density of states is smaller,
because tunneling amplitudes are larger\cite{tunme} for 
states with dominant s-wave character.
Eq.(~\ref{julliere}) is the Julli{\` e}re\cite{julliere} formula which
is in good agreement with many experimental results; 
the present derivation demonstrates that its approximate 
validity rests on the factorizability of tunneling matrix elements. 

\begin{figure}[b]
\epsfxsize=2.4in
\centerline{\epsffile{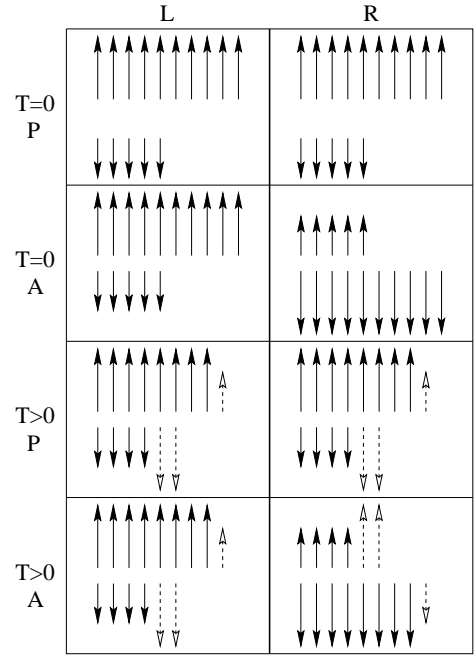}}

\vspace*{.3cm}

\caption{Schematic illustration of spin-polarized tunneling between 
left (L) and right (R) ferromagnets with parallel (P)
and antiparallel (A) magnetizations at zero and finite temperature.
In each case long arrows represent quasiparticles whose spins are 
aligned with the fluctuating bands while short arrows represent 
quasiparticles with the opposing orientation. 
The number of large arrows for $T=0$ 
is proportional to $t_a N_a$, and the number of small arrows
proportional to $t_i N_i$.  In this illustration, $t_a N_a =10$ and 
$t_i N_i =5$ in arbitrary units.  At $T=0$, 
$G = 10 \times 10 + 5 \times 5 =125$ for parallel orientations,
$G= 2 \times 10 \times 5 = 100$, for antiparallel orientations, and $MR=20\%$.
The finite temperature panels correspond to
$m(T)/m_0 = 0.6$, so that 80\% of the spectral weight resides 
in quasiparticle peaks (solid lines) and 20\% resides in shadow-band peaks
(dashed lines). The parallel orientation conductance is then reduced to
$G = 9 \times 9 + 6 \times 6 = 117$ 
,the antiparallel orientation conductance increased
to $G= 2 \times 9 \times 6 = 108$, and $MR$ decreased to $7.7\%$.} 
\label{f1}
\end{figure}

We now address the importance for the MR of 
the modifications of band-theory which are required at finite temperatures.
Up to room temperature and beyond, the main effect of 
finite $T$ on late transition element ferromagnets 
is to excite low-energy long-wavelength spin-waves.  To a good
approximation, the quasiparticle bands of the ferromagnet
adiabatically follow the local instantaneous orientation 
of the magnetic moment.  Although the quasiparticle spectrum
in this approximation is rigid, its projection onto 
the time-averaged ordered moment direction is altered.  
A quasiparticle whose spin is aligned with the fluctuating bands 
will at finite temperature be a majority spin with probability  
$(1+m(T)/m_0)/2$ and a minority spin with with probability 
$(1-m(T)/m_0)/2$.  Here $m(T)/m_0$ is the factor by which the
saturation moment is reduced at temperature $T$ due to  
thermally excited spin-waves.  A {\it shadow-band} peak must appear in
the minority-spin spectral function at the majority-spin quasiparticle
energy.  The effect on MR  of shadow-band features in the majority
and minority-spin spectral 
functions is illustrated schematically in Fig.~\ref{f1}.
Two terms contribute to $t_{\uparrow}N_{\uparrow}$ at finite temperatures, 
one arising from the majority spin quasiparticle band,
the other from the minority spin shadow band. For example, in the left-ferromagnet
\begin{eqnarray}
t_{\uparrow}^LN_{\uparrow}^L&=&\frac{1+m(T)/m_0}{2} t_a^LN_a^L+
\frac{1-m(T)/m_0}{2} t_i^LN_i^L \nonumber \\
t_{\downarrow}^LN_{\downarrow}^L&=&\frac{1+m(T)/m_0}{2}
t_i^LN_i^L+ 
\frac{1-m(T)/m_0}{2} t_a^LN_a^L \, .
\end{eqnarray}
Corresponding expressions apply for the right ferromagnet.  It follows that 
the Julli{\` e}re formula continues to apply at finite temperatures with 
a reduced polarization factor $P(T)= [m(T)/m_0] \times P(0)$ for 
each ferromagnet.  The saturation moments, $m(T)$, of bulk late transition element
ferromagnets are reduced by $\sim 5 \%$ between $T=0$
and room temperature, reduction factors are expected to 
be twice as large near the surface\cite{mills}, and will be 
larger throughout the volume of the thin film electrodes used in MTJ's. 
Reductions in $MR$ exceeding $\sim 20 \%$ at room temperature are therefore 
readily accounted for by this mechanism.
It seems likely that this effect is responsible for the 
largest part of the MR reduction seen in experiment.  An experimentally testable
prediction which follows from this assertion is that\cite{anisocaveat} 
MR should follow a $T^{3/2}$ law at low temperatures:
$MR(T) = MR(T=0)(1-A T^{3/2} + \ldots )$   
where $A$ is approximately the sum of the $T^{3/2}$ coefficients 
for the relative magnetization of the two ferromagnetic electrodes.  

We now turn to a more microscopic justification for the 
proposed shadowband features in the electron spectral 
function.  The abridged discussion presented here is intended to 
emphasize that the effect requires only that the ground state
of the ferromagnet be a spontaneously spin-polarized Fermi-liquid.
It is compatible with strong local correlation features\cite{photonew}
being present in the spectral functions, compatible  
with strong band-dependence for the quasiparticle energy spin-splitting,
and independent of the complexity and individuality of the quasiparticle
bands of particular metallic ferromagnets.  
We assume that in the ground state of the ferromagnet, the 
one-particle Green's function is that of a spin-polarized 
Fermi liquid, with quasiparticle-peaks for both $\uparrow$ and 
$\downarrow$ spins which become arbitrarily sharp as the 
Fermi energy is approached:
\begin{eqnarray} 
[G_{n\sigma,n'\sigma'}(\vec{k},E)]^{-1} &=& [\delta_{n\sigma,n'\sigma'}
E - H^{QP}_{n\sigma,n'\sigma'}(\vec{k},E)]^{-1}  \nonumber \\  
 &=&[\delta_{n\sigma,n'\sigma'} (E - E_{n,\sigma}(\vec{k},E))
 + \nonumber \\ 
 & &
 i \delta_{\sigma,\sigma'}\Gamma_{n\sigma,n'\sigma}(\vec{k},E)]^{-1}.
\label{greensfunction}
\end{eqnarray} 
The real part of the electronic self-energy is included 
in $E_{n,\sigma}(\vec{k},E)$, and we use a 
representation where the Green's function is diagonal in the band index
$n$.  Invariance of the system under spin-rotations about the 
moment direction guarantees that $G$ is diagonal in spin-indices.  At the 
Fermi energy, chosen as $E=0$, the imaginary part of the self-energy,
$\Gamma_{n\sigma,n'\sigma'}$, vanishes.  In the most naive  
version of ferromagnetic band theory the quasiparticle bands are 
rigidly spin-split, 
$E_{n,\uparrow}(\vec{k},E)\to E_n(\vec{k}) - 
\Delta/2$ and $E_{n,\downarrow}(\vec{k},E)\to E_n(\vec{k}) +
\Delta/2$.
Modern spin-density functional ferromagnet 
quasiparticle bands are {\em not} rigidly spin-split, 
and at $T=0$, are in rough\cite{dHvArefs,stoner} agreement with experiment. 

In Fe, Ni, Co and their alloys it is a good approximation, up to room
temperature and beyond, to account for  
thermal fluctuations only in the ordered moment direction,
to assume that these vary slowly on an atomic length scale, 
and to regard the quasiparticles as separate but coupled
degrees of freedom.  Keeping only the leading term in a gradient 
expansion, the excess energy density is $\rho_s | \nabla \hat n |^2 /2$
where $\hat n = (n_x, n_y, n_z) $ is the unit vector
which defines the local ordered moment
orientation and $\rho_s$ is the spin-stiffness.
For small tilts away from the $\hat z$ 
direction, the excess energy due to spin fluctuations, $H^{SW}$, 
may be expressed in terms of  
transverse components of the moment orientation vector. 
Fourier transforming we find that 
\begin{equation}
H^{SW} = \frac{\rho_s}{2 V} \sum_{\vec{q}} q^2 
\big[ n_{x - \vec{q}} \, n_{x \vec{q}} + n_{y - \vec{q}} \, n_{y \vec{q}} \big].
\label{hsw}
\end{equation}
The quantum commutator between transverse spin-components
may be replaced by a c-number as long as the saturation moment
suppression is not large:  
\begin{equation}
[n_{x - \vec{q}}, n_{y \vec{q}'} ] = i \delta_{\vec{q}, \vec{q}'} 
\frac{ 2 V} {m_0} .
\label{commutator}
\end{equation}
In Eq.(~\ref{commutator}), which results from coarse graining electron spin
commutation relations, $V$ is the system volume.
This leads to a free boson spin-wave Hamiltonian:
\begin{equation}
H^{SW} = E_0 + \sum_{\vec{q}} \epsilon_{SW}(\vec{q}) \, a^{\dagger}_{\vec{q}}
\, a_{\vec{q}}, 
\label{hswb}
\end{equation} 
where $[a_{\vec{q}}, a^{\dagger}_{\vec{q}'}] = \delta_{\vec{q}, \vec{q}'}$,
and $\epsilon_{SW}(\vec{q}) = D q^2 = 2 \rho_s q^2 / m$.
The expression we use below for spin-wave quasiparticle coupling
follows from the relationship between the spin-wave  
annihilation operator and the Fourier transforms of the transverse 
spin-orientation field:
\begin{equation}
a_{\vec{q}} = \big( \frac{m_0}{4 V} \big)^{1/2} (n_{x \vec{q}} 
+ i n_{y \vec{q}} ) 
\label{annihilation} 
\end{equation} 

In the fluctuating band picture of itinerant electron ferromagnets,
quasiparticle bands rotate in spin-space along with a slow
variation in the ordered moment direction.  At the Fermi energy 
the imaginary part of the quasiparticle self-energy can be neglected 
and 
\begin{eqnarray}
H^{QP}_{n\sigma,n'\sigma'}(\vec{k},E = 0)  &\to & 
\frac{\delta_{n,n'}}{2} ( E_{n,\uparrow}(\vec{k}) +
E_{n,\downarrow}(\vec{k} ) - 
\nonumber \\
& &
\Delta_{n}(\vec{k}) \, 
\vec {\bf \sigma}_{\sigma,\sigma'} \cdot \hat n ) 
\label{rotatebands}
\end{eqnarray} 
where $\vec{\sigma}$ are the Pauli matrices,
$\Delta_{n}(\vec{k}) \equiv  E_{n,\downarrow}(\vec{k}) -
E_{n,\uparrow}(\vec{k}) $ is the band and wavevector dependent 
spin-splitting, and all self-energies have been evaluated at 
$E=0$.  Expanding the right hand side of Eq.(~\ref{rotatebands}) 
to leading order in $n_x$ and $n_y$, the zeroth order term gives 
the ground state quasiparticle bands and the leading order 
correction gives the spin-wave quasiparticle interaction.
Allowing the orientation to vary slowly\cite{caveats} in space and Fourier 
transforming leads to the following 
interaction Hamiltonian between quasiparticles and spin-waves,
written in second quantized form: 
\begin{equation} 
H^{SW-QP} = \sum_{n,\vec{k}, \vec{q}} \frac{\Delta_{n}(\vec{k})}{V m } \, 
[ c^{\dagger}_{n,\vec{k} + \vec{q}/2,\uparrow}
c_{n,\vec{k} - \vec{q}/2,\downarrow} a^{\dagger}_{\vec{q}} + h.c. ] 
\label{couplingham}
\end{equation} 

The evaluation of single spin-wave self-energy diagrams 
corresponding to $H^{SW-QP}$ 
is similar to that for phonon-exchange\cite{mahan} in metals and leads to: 
\begin{eqnarray}
\Sigma_{n\uparrow}(\vec{k},E) &=& 
\int \frac{d \vec{q}}{(2 \pi)^3} 
{\Delta_n^2(\vec{k}+\vec{q}/2)} \times
\nonumber \\
& &
\frac{ n_B(\epsilon_{SW}(\vec{q})) + n_F(E_{n\downarrow}(\vec{k}
+ \vec{q}))} 
{[ E + \epsilon_{SW}(\vec{q}) - E_{n\downarrow} (\vec{k} + \vec{q})]}.
\label{selfenergy}
\end{eqnarray} 

Because of the gapless spin-wave excitations,
the Fermi occupation factor, $n_F$, in the 
numerator of the integrand can be neglected at finite temperatures 
in comparison with the Bose factor, $n_B$.  The integration is dominated
by the $k$-space volume satisfying $\epsilon_{SW}(\vec{q}) < k_B T$. 
We assume that at temperatures of interest  
$q's$ satisfying this inequality are small compared to Brillouin-zone
dimensions and set $\vec{q} \to 0 $ on the right hand side of 
Eq.(~\ref{selfenergy}).  These approximations lead to 
\begin{equation} 
\Sigma_{n\uparrow}(\vec{k},E) \approx \frac{\Delta_n^2(\vec{k})}{2}
\frac{1 - m(T)/m_0}{E- E_{n\downarrow}(\vec{k})}.
\label{selfenergyapprox}
\end{equation} 
where 
\begin{equation}
\frac{m(T)}{m_0} = 1 - \frac{2}{m_0} \int \frac{ d \vec{q}}{(2 \pi)^3} 
n_B(\epsilon_{SW}(\vec{q})) 
\label{magn}
\end{equation}
is the relative magnetization suppression due to thermal spin-wave excitations.  
A similar calculation gives the contribution to the 
spin $\downarrow$ self-energy:  
\begin{equation} 
\Sigma_{n\downarrow}(\vec{k},E) \approx \frac{\Delta_n^2(\vec{k})}{2}
\frac{1 - m(T)/m_0}{E- E_{n\uparrow}(\vec{k})}.
\label{selfenergyapproxb}
\end{equation} 

These simple self-energy expressions give Dyson equations 
which can be solved analytically.  The spin $\uparrow$ 
Green's function has two simple poles, one at $E_{n\uparrow}(\vec{k})$
and one at $E_{n\downarrow}(\vec{k})$.  When the magnetization 
suppression is small, the spectral weights of the two-poles
are $(1 + m(T)/m_0)/2$ and $(1-m(T)/m_0)/2$ respectively, in 
agreement with the naive fluctuating band argument given above.
Similarly the  spin $\downarrow$ Green's function splits its 
spectral weight between poles at the $T=0$ minority and 
majority band positions. 
When the magnetization suppression is large, the low-temperature
approximations leading to this result begin to fail and 
alteration of the low-temperature spectral weights will 
become more complex and more material specific.   

Experimental progress continues to   
reduce the influence of unintended extrinsic effects on the 
properties of MTJ's.
We have suggested that the temperature-dependence 
of the MR of MTJ's is a consequence of the formation of sharp shadowband
features in the one-particle Green's function 
tunneling density of states.  For a given band and spin
index, shadow bands  are located near the
quasiparticle peak of the opposite spin.  The occurrence
of shadowbands of this type is expected in the 
fluctuating band picture of metallic ferromagnets.
The temperature-dependence explanation in terms 
of intrinsic properties of bulk ferromagnets offered here, 
encourages the hope that MTJ tunneling spectroscopy 
can be developed into a probe of the
single-particle Green's function of itinerant electron 
ferromagnets with a power similar to that of photoemission.  

The authors acknowledge helpful interactions with 
W.H. Butler, J.F. Cooke, J.S. Moodera, and I. Schuller
R.S. Sooryakumar.  This work was supported by the National Science
Foundation under grants DMR-9714055 and INT-9602140 and 
by the Ministry of Education of the Czech Republic under grant ME-104.


\begin{references}
\bibitem{anisocaveat} Below $T \sim 10K$ corrections 
due to the gap in the spin-wave spectrum created by 
anisotropy and internal fields can alter this result.
\end{references}
\end{document}